\documentclass[aps,prc,twocolumn,amsmath,amssymb,superscriptaddress,showpacs,showkeys]{revtex4-2}
\usepackage{dcolumn}
\usepackage{graphicx,color}
\usepackage{multirow}
\usepackage{textcomp}
\usepackage{physics}
\usepackage{dsfont}
\usepackage{relsize}

\usepackage{tikz}	
\begin{document}
	\newcommand {\nc} {\newcommand}
	\nc {\beq} {\begin{eqnarray}}
		\nc {\eeq} {\nonumber \end{eqnarray}}
	\nc {\eeqn}[1] {\label {#1} \end{eqnarray}}
\nc {\eol} {\nonumber \\}
\nc {\eoln}[1] {\label {#1} \\}
\nc {\ve} [1] {\mbox{\boldmath $#1$}}
\nc {\ves} [1] {\mbox{\boldmath ${\scriptstyle #1}$}}
\nc {\mrm} [1] {\mathrm{#1}}
\nc {\half} {\mbox{$\frac{1}{2}$}}
\nc {\thal} {\mbox{$\frac{3}{2}$}}
\nc {\fial} {\mbox{$\frac{5}{2}$}}
\nc {\la} {\mbox{$\langle$}}
\nc {\ra} {\mbox{$\rangle$}}
\nc {\etal} {\emph{et al.}}
\nc {\eq} [1] {(\ref{#1})}
\nc {\Eq} [1] {Eq.~(\ref{#1})}
\nc {\Refc} [2] {Refs.~\cite[#1]{#2}}
\nc {\Sec} [1] {Sec.~\ref{#1}}
\nc {\chap} [1] {Chapter~\ref{#1}}
\nc {\anx} [1] {Appendix~\ref{#1}}
\nc {\tbl} [1] {Table~\ref{#1}}
\nc {\Fig} [1] {Fig.~\ref{#1}}
\nc {\ex} [1] {$^{#1}$}
\nc {\Sch} {Schr\"odinger }
\nc {\flim} [2] {\mathop{\longrightarrow}\limits_{{#1}\rightarrow{#2}}}
\nc {\textdegr}{$^{\circ}$}
\nc {\inred} [1]{\textcolor{red}{#1}}
\nc {\inblue} [1]{\textcolor{blue}{#1}}
\nc {\IR} [1]{\textcolor{red}{#1}}
\nc {\IB} [1]{\textcolor{blue}{#1}}
\nc{\pderiv}[2]{\cfrac{\partial #1}{\partial #2}}
\nc{\deriv}[2]{\cfrac{d#1}{d#2}}
\title{Green's function knockout formalism}
\author{C.~Hebborn}
\email{hebborn@frib.msu.edu}
\affiliation{Facility for Rare Isotope Beams, Michigan State University, East Lansing, MI 48824, USA}
\affiliation{Lawrence Livermore National Laboratory, P.O. Box 808, L-414, Livermore, California 94551, USA}
\email{hebborn@frib.msu.edu}
\author{G.~Potel}
\email{potelaguilar1@llnl.gov}
\affiliation{Lawrence Livermore National Laboratory, P.O. Box 808, L-414, Livermore, California 94551, USA}

\date{\today}
\begin{abstract}
	Knockout nuclear   reactions, in which a nucleon is removed from a nucleus as a result of the collision with another nucleus, have been widely used as an experimental tool, both to populate isotopes further removed from stability, and to obtain information about the single-particle nature of the nuclear spectrum. In order to fully exploit the experimental information, theory is needed for the description of both the structure of the nuclei involved, and the dynamics associated with the nucleon removal mechanisms.  The standard approach, using  theoretical shell-model spectroscopic factors for the structure description coupled with an eikonal model of reaction,   has been  successful when used in the context of the removal of valence nucleons in nuclei close to stability. However, it has been argued that the reaction theory might need to be revisited in the case of exotic nuclei, more specifically for highly asymmetric nuclei in which the deficient species (neutrons or protons) is being removed. We present here a new formalism for the nucleon-removal and -addition reaction through knockout and transfer reactions, that treats consistently structure and reaction  properties using dispersive optical potentials.  In particular, our formalism includes the dynamical effects associated with the removal of a {nucleon} from the projectile, which  might  explain the long standing puzzle of the quenching of spectroscopic factors in nuclei with extreme neutrons-to-protons ratios. 
	
\end{abstract}

\maketitle
%


\section{Introduction}\label{Introduction}

Radioactive Ion Beams (RIBs) facilities are transforming the field of low-energy nuclear physics by setting short-lived, exotic isotopes within experimental reach. The  availability of new experimental data has been matched by theoretical efforts towards the description of nuclear systems away from the stability valley, and, more generally, towards an understanding of nuclear structure in an exotic context \cite{Oetal20}.  The corresponding paradigm shift in the theory of nuclear structure has to  be complemented by a revision of nuclear reaction theory, needed for the description of the experiments in which radioactive ions are involved. Within this context, a considerable effort has been devoted recently to the description of reactions with weakly-bound {nuclei~(see, e.g. \cite{HAGINO2022103951,BC12,Bonaccorso18,BB22}, see also \cite{Nunes:20} and references therein)}.

However, the study of neutron-rich (resp. proton-rich)  raises a complementary question about the behaviour of the deeply-bound protons (resp. neutrons) belonging to the same nucleus. This question has been highlighted in  publications by Gade and collaborators~\cite{Gade:08,TG14,TG21}, in which they present a review of results of one-neutron (resp. one-proton) knockout experiments expressed as a function of the difference $\Delta S=S_n-S_p$ between the neutron $S_n$ and proton separation  $S_p$ energies (resp. $\Delta S=S_p-S_n$). In this work, they arrive at the puzzling conclusion that theory is unable to account for as much as 80\% of the quenching of single-particle strength when the knocked out particle belongs to the deficient species in systems with a large value of $\Delta S$. Several authors have suggested that nuclear structure calculations might fail to fully  account for short-range correlations between neutrons and protons in  highly asymmetric systems, leading to an overestimation of the single-particle content of the states populated in knockout reactions~\cite{Jetal11,PhysRevC.104.L061301}.

On the other hand, the fact that the strong dependence on $\Delta S$ of the spectroscopic factor quenching is not observed in transfer and quasifree scattering experiments  has led some authors to suggest that the issue might be in the  theory associated with the description of the reaction process   in the case of knockout experiments~\cite{Tetal09,Letal10,Fetal12,Fetal13,GM18} (see Ref.~\cite{Aetal21} for a recent review). Knockout experiments are often described within the eikonal model~\cite{G59,HT03}, assumed to be valid for high beam energies. The sudden {and the core spectator}  approximations, in which the nuclear degrees of freedom are frozen during the collision process, are  used to describe the one-nucleon removal from the projectile. The core spectator approximation is based on the asumption that the characteristic decay times of the core states populated in the nucleon removal process are large compared to the collision time. This seems reasonable when the nucleon is removed from a state not too far away from the Fermi energy on a stable nucleus, since the narrow associated energy width is  small, the corresponding damping (decay) time being therefore rather large.

However, it has been well known since  knockout experiments were performed in the early 60's (see, e.g., \cite{Jacob:66}) that  hole states associated with the removal of deeply-bound nucleons have much larger widths, sometimes of the order of tens of MeV. In other words, hole states resulting from the removal of a deeply-bound nucleon can decay into complicated many-body nuclear states rather quickly, in times of the order of $10^{-23}$-$10^{-22}$~s, which is comparable to the short collision times associated with fast knockout experiments. This damping process results in dissipative effects associated with the dynamics of hole states in nuclei, not taken into account within the  core spectator approximation. These effects have already been estimated in Ref.~\cite{Letal11} within the intra-nuclear cascade approach, showing that they can excite the core above the particle emission threshold, leading to particle evaporation and a loss of flux in the outgoing channel that could account for the observed spectroscopic factor quenching. {This discussion applies to the cases in which the residual core is the only species detected in the experiment, such as in most standard knockout reactions.} 

The above parlance highlights the general need to integrate in a consistent theoretical framework the structure and dynamics of  many nucleons in a nucleus, and the description of  reactions used to study them in an experimental context. An attempt to account for the  dynamics between a nucleus and a transferred nucleon has already been implemented in the the Green's Function Transfer (GFT) formalism, albeit in a  one-nucleon addition context~\cite{PotelNunesThompson,Petal17} (see also equivalent theories in Refs. \cite{LM15a,LM15b,CCS16}). We present here an extension of this idea, the Green's Function Knockout (GFK), which describes   one-nucleon removal processes.

In Sec. \ref{SecGFK}, we derive the general formalism for one-nucleon removal reactions, such as knockout and $(p,d)$, and we show in Sec. \ref{S3} its connection to one-neutron addition processes and the GFT formalism. In Sec. \ref{SecDistorWF} we discuss  different approximations that can be made in the context of the GFK, before concluding in Sec. \ref{Conclusions} with a summary and outlook of future developments.

\section{Green's function knockout formalism }\label{SecGFK}
Let us introduce the GFK by considering a   reaction involving a projectile nucleus $P$ of mass $m_P$ impinging on a target $T$ of mass $m_T$.  This system is described by the solution $\Psi$ of the many-body Hamiltonian $\hat H$
\begin{align}
\label{eq:0}
\hat H \ket{\Psi}=E \ket {\Psi},
\end{align}
where  $E$ is the  energy of the system in the center of mass frame.
We are interested in the reaction  channel in which a nucleon $N$ is knocked out from the projectile  and only the  residual core $c$ is detected  with a kinetic energy $E_{f_{cT}}$. Since the final state of the $N$-$T$ system is  often not measured,  we focus here on deriving the inclusive cross section in both the core $c$ and $N$-$T$ systems, i.e.,  summed over all energy-conserving final states $f_{{NT}}$ of the $N$-$T$ system and   over all the  energetically available states $f_c$ of the core $c$ \cite{HM85,HT03,IAV85}. When expressed as a function of the deflection
angle of the core $\Omega$ and its final kinetic {energy} $E_{f_{cT}}$, the cross section in the {\it prior} representation reads
\begin{align}
\label{eq:1}
\nonumber &\frac{d\sigma }{dE_{f_{cT}}d\Omega}=\frac{2\pi\mu_{PT}}{\hbar^2 k_{PT}}\,\rho(E_{f_{cT}})\\
&\nonumber\times\sum_{f_{NT},f_c}\left|\left\langle\left.\Psi(f_{NT},f_c)\,\right|\hat V_{prior}|\phi_T^{(0)}\phi_P^{(0)}\mathcal{F}\right\rangle\right|^2\\
&\times\delta(E-E_{f_{cT}}-E_{f_c}-E_{f_{NT}}),
\end{align}
where $\Psi(f_{NT},f_c)$ is the solution of (\ref{eq:0}) subject to the boundary condition consisting in having an outgoing wave containing the core in a state $f_c$ moving with kinetic energy $E_{f_{cT}}$ with respect to the target, and the $N$-$T$ system is in a state $f_{NT}$.  The many-body wave functions $\phi_T^{(0)}$   and $\phi_P^{(0)}$ correspond to the ground state of the target and the projectile, respectively, while $\mathcal{F}$ describes the relative motion between the projectile and the target in the incoming channel. 
The term $\rho(E_{f_{cT}})=\mu_{cT}k^f_{cT}/\left[(2\pi)^3\hbar^2\right]$ is the asymptotic phase space factor (density of states) of the core fragment,  $k^f_{cT}$ and $k_{PT}$ are  respectively the final $c$-$T$ and initial $P$-$T$ wave numbers, $\mu_{cT}$  and $\mu_{PT}$ {being} the $c$-$T$ and  $P$-$T$ reduced masses.

\begin{figure}
 	\centering
 	\begin{tikzpicture}[scale=0.7]
 	\large
 	\begin{scope}[>=latex]
 	\filldraw[color=black!40!white] (10,-2) circle (0.5cm);
 	\draw (10,-2)++(0.5,0) node[anchor=north west] {$T$};
 	\filldraw[color=black!40!white] (0,1) circle (0.45cm);
 	\filldraw[color=black!40!white] ({sqrt(2)*3/2},{1+sqrt(2)*3/2}) circle (5pt);
 	\draw ({sqrt(2)*3/2},{1+sqrt(2)*3/2}) node[anchor=south west]{$N$};
 	\draw (-0.45,{1-0.45}) node[anchor= east]{$c$};
 	\draw[dashed, shift={({sqrt(2)/2+0.2},{1.2+sqrt(2)/2})}, rotate=45] (0,0) ellipse (3cm and 1.5cm);
 	\draw ({sqrt(2)*3/2},{2+sqrt(2)*3/2}) node[anchor=south west](xx){$ $};
 	\draw ({sqrt(2)/2},{1+(sqrt(2)/2)})++(-1.25,1.25) node[anchor=south east]{$P$};
 	\draw ({sqrt(2)*3/2},{2+sqrt(2)*3/2}) node[anchor=south west](xx){$ $};
 	\draw[->] (0,1) -- ({sqrt(2)*3/2},{1+sqrt(2)*3/2});
 	\filldraw (0,1)++({sqrt(2)/2},{(sqrt(2)/2)}) circle (1pt);
 	\draw[<-] ({sqrt(2)/2},{1+(sqrt(2)/2)}) -- ({(sqrt(2)/4+10/2)},{(1+(sqrt(2)/2))/2-2/2});
 	\draw (5.3303,0.07)  node[anchor=south east]{$\ve{R}$};
 	\draw ({(sqrt(2)/4+10/2)},{(1+(sqrt(2)/2))/2-2/2}) -- (10,-2);
 	\draw[->] (10,-2) -- (0,1);
 	\draw (5,-0.5)  node[anchor=north east]{$\ve{R_{cT}}$};
 	\draw[->] (10,-2) -- ({sqrt(2)*3/2},{1+sqrt(2)*3/2});
 	\draw (7.4607,0.5607) node[anchor=south east]{$\ve{R_{NT}}$};
 	\draw (0,1)++({sqrt(2)+0.2},{sqrt(2)}) node[anchor=south east]{$\ve{r}$};
 	\draw[->] ({sqrt(2)/2},{1+(sqrt(2)/2)}) -- ({sqrt(2)/2+2.3},{1+(sqrt(2)/2)}) node[anchor=north west] {$\ve{k}_{PT}$};
 	\draw[dotted] (0,-2)-- (9.5,-2);
 	\draw[->]  (1,-2) -- (3,-2) node[anchor=north] {$\ve{\widehat{Z}}$};
 	\end{scope}
 	\end{tikzpicture}
 	\caption{Set of coordinates used in this article: the $c$-$N$, $c$-$T$ and $N$-$T$ relative coordinates $\ve r$, $\ve R_{cT}$ and $\ve R_{NT}$, respectively.}\label{Fig1}
  \end{figure}
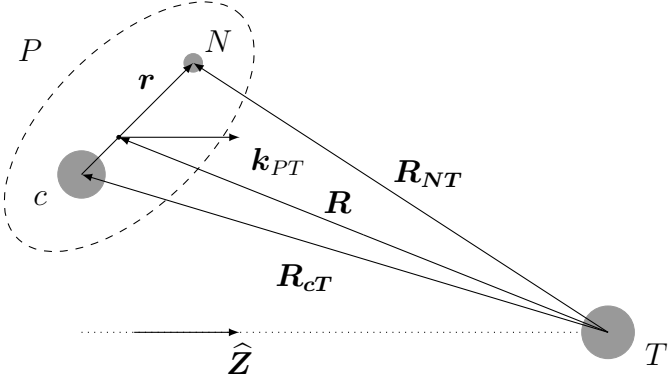

In its exact form, the prior potential $\hat V_{prior}$ depends on the $c$-$T$ and $N$-$T$ many-body potentials, respectively $\hat V_{cT}$ and $\hat V_{NT}$, and on the  potential  $\hat V_i$ used to compute the incoming scattering function $\mathcal{F}$
\begin{eqnarray}
\hat V_{prior}=\hat V_{cT}+\hat V_{NT}-\hat V_{i}.\label{eq300}
\end{eqnarray}
 The exact prior potential therefore depends on  the $N$-$T$ and $c$-$T$  relative coordinates, respectively  $\ve R_{NT}$ and $\ve R_{cT}$ (see Fig. \ref{Fig1}), and the intrinsic coordinates of the target and the core, respectively   $\ve \xi_T$ and $\ve \xi_c$.
In this formalism, we neglect the dependence of the prior potential on the core intrinsic coordinates $\ve \xi_c$, although the role of reaction channels in exciting the core is accounted for with the inclusion of an imaginary part. 
In Sec. \ref{SecDistorWF}, we will discuss different possible choices of $\mathcal{F}$ and $\hat V_i$.

We now approximate the exact wave function as
\begin{equation}
  \label{eq:200}
   \Psi(f_{NT},f_c)\approx\chi_{cT}^{(f_{cT})}(\ve R_{cT})\psi_{NT}^{(f_{NT})}(\ve R_{NT},\ve \xi_{T})\psi_{c}^{(f_c)}(\ve \xi_{c})
\end{equation}
where $\chi_{cT}^{(f_{cT})}$ is the  wave function describing the  final $c$-$T$ relative motion, and the final wave functions  of the core and the $N$-$T$  systems, respectively  $\psi^{(f_c)}_c$ and $\psi_{NT}^{(f_{NT})}$,  are many-body objects.  
These functions satisfy the Schr\"odinger equations
\begin{align}
\label{eq:2}
  &\left(E_{f_c}-\hat h_c\right)\,\psi_{c}^{(f_c)}(\ve \xi_c)=0,\\
&\left(E_{f_{NT}}-\hat T_{NT}-\hat V_{NT}-\hat h_T\right)\,\psi_{NT}^{(f_{NT})}(\ve R_{NT},\ve \xi_T)=0, \label{eq:4}
\end{align}
where we define   the many-body core ($\hat h_c$) and target ($\hat h_T$) Hamiltonian operators,   and the kinetic energy  operator $\hat T_{NT}$. Then,
\begin{align}
\label{eq:201}
\nonumber &\frac{d\sigma }{dE_{f_{cT}}d\Omega}=\frac{2\pi\mu_{PT}}{\hbar^2 k_{PT}}\,\rho(E_{f_{cT}})\\
&\nonumber\times\sum_{f_{NT},f_c}\left|\left\langle\chi_{cT}^{(f_{cT})}\,\psi_{NT}^{(f_{NT})}\psi_{NT}^{(f_c)}\,\right| \hat V_{prior} \left|\phi_T^{(0)}\phi_P^{(0)}\mathcal{F}\right\rangle\right|^2\\
&\times\delta(E-E_{f_{cT}}-E_{f_c}-E_{f_{NT}}).
\end{align}
  
We now write the sum over delta functions in Eq.~\eqref{eq:201}, which enforces energy conservation, in terms of the imaginary part of the Green's function~\cite{Dickhoff:05}
\begin{align}
\label{eq:5}
\nonumber \sum_{f_{NT},f_c}&\left|\psi_{NT}^{(f_{NT})}\psi_{c}^{(f_{c})}\right\rangle\left\langle\psi_{c}^{(f_{c})}\psi_{NT}^{(f_{NT})}\right|\\
&\hspace{-1cm}\times \delta(E-E_{f_{cT}}-E_{f_c}-E_{f_{NT}})=-\frac{1}{\pi}\text{Im }\hat G(E-E_{f_{cT}}),
\end{align}
which expresses the relationship between the spectral function and the many-body Green's function
\begin{align}
\label{eq:6}  \hat G(\mathcal {E})&=\lim_{\eta\to0}\frac{1}{\mathcal {E}-\hat h_c-\hat T_{NT}-\hat V_{NT}-\hat h_{T}+i\eta}.
\end{align}
Making use of the complete set of product states $\psi_{NT}^{(f_{NT})}$ and $\psi_c^{(f_c)}$, the Green's function can be written in the Lehmann representation, 
\begin{align}
	\label{eq6b}
	& \hat G(\mathcal {E})=\lim_{\eta\to0}\sum_{f_{NT},f_c}\frac{\left|\psi_{NT}^{(f_{NT})}\psi_{c}^{(f_{c})}\right\rangle\left\langle\psi_{c}^{(f_{c})}\psi_{NT}^{(f_{NT})}\right|}{\mathcal {E}-E_{f_c}-E_{f_{NT}}+i\eta},  
\end{align}
or in two other equivalent  representations,
\begin{align}
\label{eq:7}
  & \hat G_{h}(\mathcal {E})=\lim_{\eta\to0}\sum_{f_{NT}}\frac{\left|\psi_{NT}^{(f_{NT})}\right\rangle\left\langle\psi_{NT}^{(f_{NT})}\right|}{\mathcal {E}-E_{f_{NT}}-\hat h_c+i\eta},  \\
&\hat G_{NT}(\mathcal {E})=\nonumber \\\label{eq:8}
&\lim_{\eta\to0}\sum_{f_{c}}\frac{\left|\psi_{c}^{(f_{c})}\right\rangle\left\langle\psi_{c}^{(f_{c})}\right|}{\mathcal {E}-E_{f_c}-\hat T_{NT}-\hat V_{NT}-\hat h_T+i\eta}.    
\end{align}

Using   \Eq{eq6b}  we can rewrite (\ref{eq:201}) as
\begin{align}\label{eq:11b}
&\frac{d\sigma }{dE_{f_{cT}}d\Omega}=-\frac{2\mu_{PT}}{\hbar^2 k_{PT}}\,\rho(E_{f_{cT}})\nonumber\\
\nonumber &\times \left\langle\phi_T^{(0)}\phi_P^{(0)}\mathcal{F}\,\right|\hat V_{prior} \left|\,\chi^{(f_{cT})}_{cT}\right\rangle \\
  &\times\text{Im }\hat G(E-E_{f_{cT}})  \left\langle\chi^{(f_{cT})}_{cT}\,\right| \hat V_{prior} \left|\phi_T^{(0)}\phi_P^{(0)}\mathcal{F}\right\rangle,
\end{align}
or, equivalently,
\begin{align}
&\frac{d\sigma }{dE_{f_{cT}}d\Omega}=-\frac{2\mu_{PT}}{\hbar^2 k_{PT}}\,\rho(E_{f_{cT}})\nonumber\\
\nonumber &\times \left\langle\phi_T^{(0)}\phi_P^{(0)}\mathcal{F}\,\right| \hat V_{prior} \left|\,\chi^{(f_{cT})}_{cT}\right\rangle \\
  &\times\text{Im }\hat G_h(E-E_{f_{cT}})  \left\langle\chi^{(f_{cT})}_{cT}\,\right| \hat V_{prior} \left|\phi_T^{(0)}\phi_P^{(0)}\mathcal{F}\right\rangle.\label{eq:11}
\end{align}
This expression is still difficult to handle as it contains  the  many-body operator $\hat G_h$.

 In order to reduce the dimensionality of the problem, we average the Green's function over the ground state of the projectile
\begin{align}
\label{eq:14}
&\left\langle \phi_P^{(0)}\right|\hat G_h(E-E_{f_{cT}})\left|\phi_P^{(0)}\right\rangle\nonumber\\&=\lim_{\eta\to0}\sum_{f_{NT}}\left|\psi_{NT}^{(f_{NT})}\right\rangle\left\langle\psi_{NT}^{(f_{NT})}\right|\nonumber \\
\nonumber &\times\,\left\langle\phi_P^{(0)}\right|\frac{1}{E-E_{f_{cT}}-E_{f_{NT}}-\hat h_c+i\eta}\left|\phi_P^{(0)}\right\rangle.  \\
&=\lim_{\eta\to0}\sum_{f_{NT}}\left|\psi_{NT}^{(f_{NT})}\right\rangle\left\langle\psi_{NT}^{(f_{NT})}\right| \hat G^{opt}_{h}(E-E_{f_{cT}}-E_{f_{NT}})
\end{align}
where we have  defined the optical reduction of the Green's function 
\begin{align}
\label{eq:15}
\nonumber \hat G^{opt}_{h}(E_{h})&=\lim_{\eta\to0}\left\langle \phi_P^{(0)}\right|\left(E_h-\hat h_c+i\eta\right)^{-1}\left|\phi_P^{(0)}\right\rangle\\
&=\lim_{\eta\to0}\left(E_h-\hat T_h-\hat U_h+i\eta\right)^{-1},
\end{align}
which is a one-body operator.   In this equation,   we have defined  the one-particle $\hat T_h$ kinetic and $\hat U_h$ hole  potential  operators, which in general is non-local, complex and energy-dependent \cite{Dickhoff:19}. 
The eigenstates of the hole Hamiltonian correspond to discrete overlap functions \begin{equation}
\psi_{h}^{(f_{c})}(\ve r)= \braket{\phi_P^{(0)}}{\psi_c^{(f_c)}}, 
\end{equation}solutions to the
 Schr\"odinger equation \cite{Dickhoff:05,Dickhoff:19}
\begin{align}
&\left(E_h-\hat T_h-\hat U_h\right)\,\psi_{h}^{(f_{c})}(\ve r)=0.\label{eq:18}
\end{align}

The energy needed to promote a nucleon of the projectile $P$ to a zero-energy state, leaving the core in its ground state, is the nucleon separation energy $S_N^{(P)}$, which we define positive for particle-bound systems, according to the standard convention. Therefore, the core ground state  is obtained by creating  a hole of energy $E_h=-S_N^{(P)}$ in the projectile. An excited state of the core with energy $E_{f_c}$ is obtained by delivering the corresponding additional energy, thus creating a deeper hole, 
\begin{equation}\label{eq:100}
	E_h=-E_{f_c}-S_N^{(P)}.
\end{equation}
{Note that Eq.~\eqref{eq:18}    does  not constrain the proper normalization of $\psi_{h}^{(f_{c})}$, namely its spectroscopic factor $S$}.  Since  {a dispersive} optical potential can be identified with the self-energy of the nucleon in the nuclear medium, the  spectroscopic factor  is connected with the energy dependence of the hole potential,
\begin{equation}
\label{eq:19}
S(E_{f_c})=\left(1-\left.\frac{\partial U_h(E)}{\partial E}\right|_{E_h}\right)^{-1}.
\end{equation} 
{The above equation is verified only for dispersive potentials, and expresses the relationship between the optical potential and the energy distribution of single particle strength \cite{Dickhoff:19,Dickhoff:05}. It highlights the importance of the use of dispersive potentials in  the present formalism, where it is essential to have an accurate description of the spectral function, including the verification of sum rules based on the conservation of the number of particles.}

By also defining the transition amplitudes
\begin{widetext}
\begin{align}
\label{eq:20}
  T^{(f_{NT})}(\ve R_{NT},\ve R_{cT})=
        \int \mathcal{F}(\ve R_{cT},\ve R_{NT},\ve \xi_T)    \phi_T^{(0)}(\ve \xi_T)\,\psi^{{(f_{NT})}*}_{NT}(\ve R_{NT},\ve \xi_T) \hat V_{prior} (\ve R_{NT},\ve R_{cT},\ve \xi_T)d\ve \xi_{T},
\end{align}
\end{widetext}
and the source term
\begin{align}
\label{eq:21}
&\rho_h^{(f_{NT})}(\ve r)=\int \chi^{(f_{cT})*}_{cT}( \ve R_{cT})\, T^{(f_{NT})}(\ve R_{NT},\ve R_{cT}) \,d\ve R_{cT},  
\end{align}
Eq.~\eqref{eq:11} becomes
\begin{align}
\label{eq:22}
&\frac{d\sigma }{dE_{f_{cT}}d\Omega}=-\frac{2\mu_{PT}}{\hbar^2 k_{PT}}\,\rho(E_{f_{cT}})\nonumber \\
&\times\, \sum_{f_{NT}}\left\langle\rho_h^{(f_{NT})} \right |\text{Im }\hat G^{opt}_{h}(E_h) \left |\rho_h^{(f_{NT})}\right\rangle,
\end{align}
where the argument of the Green's function reflects energy conservation, {i.e.} 
\begin{align}\label{eq:202}
E_{f_{c}} &=E-  E_{f_{cT}}- E_{f_{NT}},
\end{align} 
{complemented with}  (see Eq. (\ref{eq:100}))
\begin{align}
\label{eq:203}
  E_h=E_{f_{cT}}+E_{f_{NT}}-E-S_N^{(P)}.
\end{align} 
Eq.~\eqref{eq:22} can be interpreted in terms of the  source term~\eqref{eq:21}   expressing the probability of  the  production of a hole in the projectile $P$ at a position $\ve r$, while the Green's function describes the dynamical evolution  of the core-hole system. Dissipative effects associated with core excitation are connected with the imaginary part of the optical potential, and are fully accounted for. When these excitations take place at energies above the nucleon emission threshold they can result in particle evaporation, and thus  reduce the  knockout cross section of deeply-bound nucleons~\cite{Letal11}.

In order to highlight  the effect of the absorption of the core by the hole, we will express Eq.~\eqref{eq:22} in terms of $\hat U_h$ by 
 first  defining the free hole propagator $\hat G_h^{opt,0}$ as 
\begin{equation}
	\label{eq:23}
\hat G_h^{opt,0}(E_h)=\lim_{\eta\to0}\left(E_h-\hat T_h+i\eta\right)^{-1},
\end{equation}
so that we can write
\begin{eqnarray}
(\hat G_h^{opt,0})^{-1}-(\hat G^{opt}_{h})^{-1}=\hat U_h.
\end{eqnarray}
By manipulating this expression, we obtain
\begin{align}
&\hat G_h^{opt,0}\left[(\hat G_h^{opt,0})^{-1}-(\hat G^{opt}_{h})^{-1}\right]\hat G^{opt}_{h}\nonumber\\
\nonumber & =\hat G^{opt}_{h}-\hat G_h^{opt,0}\\
&= \hat G_h^{opt,0}\hat U_h\hat G^{opt}_{h},
\end{align}
which leads to   the Dyson equation \begin{eqnarray}
\hat G^{opt}_{h}=\hat G_h^{opt,0}+\hat G_h^{opt,0}\hat U_h\hat G^{opt}_{h}.
\end{eqnarray}
We can now rewrite this propagator as 
\begin{align}
\label{eq:28}
  \nonumber \hat G^{opt}_{h}&=\left(1+ \hat G^{opt\dagger}_{h}\hat U_h^\dagger\right)\hat  G_h^{opt,0}\left(1+\hat U_h\hat G^{opt}_{h}\right)\\
  &-\hat G^{opt\dagger}_{h} \hat U_h^\dagger \hat G^{opt}_{h}.
\end{align}
The first term of Eq.~\eqref{eq:28}   corresponds to scattering states of the core-hole system. Since  the hole Hamiltonian describes the removal of a {bound} nucleon {from} the projectile ground state, it does not have any scattering solutions and this first term vanishes.  We therefore obtain
\begin{equation}
{\rm Im}\,\hat G^{opt}_{h}=\hat G^{opt\dagger}_{h} {\rm Im}\,\hat U_h \hat G^{opt}_{h}.	\label{eq:29}
\end{equation}

By defining the hole wave function 
\begin{align}
\label{eq:30}
\phi^{(f_{NT})}_h(\ve r)&=\hat G^{opt}_{h}(E_h)\,\rho^{(f_{NT})}_h(\ve r)
\end{align}
we can then write the  cross section~\eqref{eq:22} as
\begin{align}
\nonumber &\frac{d\sigma }{dE_{f_{cT}}d\Omega}=-\frac{2\mu_{PT}}{\hbar^2 k_{PT}}\,\rho(E_{f_{cT}}) \\
&\times\,\sum_{f_{NT}} \left\langle\phi^{(f_{NT})}_h \right |{\rm Im}\, \hat U_h(E_{h}) \left |\phi_h^{(f_{NT})}\right\rangle.
\end{align}
If we assume that for excitation energies above the first particle emission threshold $S_x^{(c)}$ the core will evaporate particles, the experimental cross section for observing the core $c$ is restricted to    $0<E_{f_c}<S_x^{(c)}$ and therefore the hole energy is restricted to $-S_x^{(c)}-S_N^{(P)}<E_h<-S_N^{(P)}$. The cross section therefore reads
\begin{widetext}
\begin{align}
\frac{d\sigma }{dE_{f_{cT}}d\Omega}=-\frac{2\mu_{PT}}{\hbar^2 k_{PT}} \,\rho(E_{f_{cT}})\sum_{E_{h}=-S_x^{(c)}-S_N^{(P)}}^{-S_N^{(P)}}  \left\langle\phi_h^{(f_{NT})} \right |{\rm Im}\, \hat U_h(E_{h}) \left |\phi_h^{(f_{NT})}\right\rangle.   \label{eq:33}
\end{align}
\end{widetext}

Let us stress  that the above expression is inclusive in both the $N$-$T$ and the final states of the core. The sum over all energy-accessible final states of the core is implied in the imaginary part of the hole potential ${\rm Im}\, \hat U_h$, while it is explicit in the $N$-$T$ channel, as  we sum over the hole states $\phi_h^{(f_{NT})}$, which contain the $N$-$T$ transition amplitudes in the source term (\ref{eq:21}). 
The only approximations  made in the derivation of Eq. (\ref{eq:33}) are the assumption that the prior potential does not depend explicitly on the core internal coordinates, and the factorization of the many-body wave function (\ref{eq:200}).  In particular, no sudden or core spectator approximations has been  made   concerning the dynamics of the $N$-$T$ and $N$-$c$ systems. This is in  contrast  with the standard eikonal framework~\cite{HT03,HM85,Gade:08}, where dynamical, non-sudden  effects associated with the nucleon extraction from the projectile (or hole creation) are neglected. It is reasonable to expect that these effects will be particularly important for the knockout of deeply-bound nucleons, creating deeply-bound holes, which could give rise to core excitation above particle emission thresholds and therefore to particle evaporation. {The quenching of spectroscopic strength is further enhanced in  systems with high neutron-proton asymmetry by the fact that the emission threshold of the deficient species tends to be low, and the sum  in Eq. (\ref{eq:33}) is severely restricted}.

\section{Application to $A(p,d)B$ and $B(d,p)A$ transfer reactions}\label{S3}
Let us now   consider a pickup reaction, in which a neutron is transferred from the projectile $A(\equiv B+n)$ to the proton target $p$, resulting in the formation of a deuteron $d$ and the core nucleus $B$. In these reactions, the deuteron  is detected   with kinetic energy $E_{f_d}$, and the residual nucleus $B$ is not observed. While this process is still inclusive in the core ($B$) channel, it is exclusive in the $n$-$p$ channel, in which the  deuteron has been detected in {the only available}  bound state, namely its ground state.  According to the notation of the previous section,	
\begin{align}
c\equiv B \quad T\equiv p \quad N\equiv n \quad c+N\equiv A\quad T+N\equiv d.
\end{align}
{In this case, the deuteron ground state, with binding energy $\epsilon_d=-2.2246$ MeV, is the only one to be kept in the sum appearing in Eq. (\ref{eq:33}).  The final $N$-$T$ wave function corresponds to the ground state of the deuteron $\psi_{NT}^{(f_{NT})}\equiv \phi_d$, and  the differential cross section becomes
\begin{equation}
\label{eq:34}
\frac{d\sigma}{dE_{f_d}d\Omega}=-\frac{2\mu_{pA}}{\hbar^2 k_{pA}}\,\rho(E_{f_d})  \left\langle\phi_h \right |{\rm Im}\, \hat U_h(E_{h} )\left |\phi_h\right\rangle,
\end{equation}
where 
$k_{pA}$ is the initial $p$-$A$ wave number, $\mu_{pA}$ is the $p$-$A$ reduced mass, $E_h={E_{f_d}}+\epsilon_d-E-S_n^{(A)}=-E_{f_B}-S_n^{(A)}$ is the  $B$-hole  energy,  ${E_{f_B}}$ {is the energy of the final state of the residual nucleus $B$ and $S_n^{(A)}$ is the neutron separation energy of the $A$ nucleus.}
The GFK formalism describes non-sudden, dissipative processes in the core $B$ system, and takes into account the quantum many-body dynamics ({encoded here in  the imaginary part of the hole optical potential, ${\rm Im}\, \hat U_h$}) to describe its final state.  {Thanks to the use of dispersive optical potentials, the  GFK formalism allows a consistent description of pickup reactions leaving the residual nucleus in a bound and resonant state.} Typically, ${\rm Im}\, \hat U_h$ will be spin- and parity-dependent. Within this context, this framework might be useful in situations in which one is interested in the determination of the energy, spin, and parity of the final state of the residual nucleus, as in, e.g., $(p,d)$ surrogate reactions~\cite{Eetal12,Eetal18}.

In reactions where a neutron is transferred from the target to the projectile, such as $B(d,p)A$, the  state of the nucleus $A$ is not measured. However, as for pickup reactions, the  energy of the state of the nucleus $A$ can be deduced from the energy of the final proton $E_{f_p}$. In this case, we have 
\begin{equation}
\label{eq:35}
\frac{d\sigma}{dE_{f_p}d\Omega}=-\frac{2\mu_{d}}{\hbar^2 k_{d}}\,\rho(E_{f_p})\left\langle\phi_n \right |{\rm Im}\, \hat  U_n( E_n )\left |\phi_n\right\rangle,
\end{equation}
where $k_d$ is the initial wave number of the $d$-$B$ system, $\mu_d $ is the $d$-$B$ reduced mass,  $E_n$ is the excitation energy  relative to the ground state of the final nucleus $A$ and $ \hat U_n$ is the  $n$-$B$ optical potential.
Note that if the incoming scattering wave function $\mathcal{F}$ is obtained within the Distorted Wave Born Approximation (DWBA) described in Sec.~\ref{DWBA}, this expression is similar to Eq. (25) of Ref.~\cite{PotelNunesThompson}, {and the definition of the $n$-$B$ wave function $\phi_n$ 
is analogous to  the one  in Sec.~\ref{SecGFK}}. The framework presented here represents therefore a generalization of the GFT~\cite{PotelNunesThompson} to describe inclusive measurements of both one-nucleon addition and removal reactions.   Compared to previous models, the GFK is therefore applicable to both transfer  and knockout reactions. This general applicability of the GFK   is a great advantage, as it provides a theoretical framework to compare their analysis and  gives insights on what are the reaction mechanisms at play.

Consequently, the GFK might allow  to  shed some light on the discrepancy between  the analyses of transfer and knockout experiments for projectile with large neutron-to-proton asymmetry~\cite{Gade:08,TG14,TG21,Tetal09,Letal10,Aetal21}. {Since in transfer reactions, the knocked-out nucleon is measured (in the case of ($p,d$), it  forms a deuteron with the proton target),  any dissipative effects associated with the extraction of the neutron from the projectile $A$, will not impact the cross sections (except if the excited residual nucleus is able to emit deuterons). In knockout reactions, since  it is the residual nucleus that is detected, the cross section is therefore directly impacted by these particle emissions of the residual nucleus. Since these effects are expected to be more important for the removal of deeply-bound nucleons, this suggests that that these dissipative effects might contribute to explain the discrepancy between the analyses of transfer and knockout data as discussed in Ref.~\cite{Letal11}. }

\section{Computation of the incoming  scattering wave}\label{SecDistorWF}

To implement the GFK formalism, it is useful to approximate   the  incoming scattering wave $\mathcal{F}$.  We present below two possible approaches, which will also determine the choice of the potential $\hat V_{prior}$ used in the transition amplitudes (\ref{eq:20}).

\subsection{Eikonal approximation}
Reactions measured at  energies above 60~MeV/nucleon, such as breakup reactions, are accurately described by the eikonal model~\cite{HT03,BCG05,BCDS09,PDB12,HC21}. This approximation~\cite{G59} assumes that the projectile-target relative motion does not differ much from the initial plane wave $\chi_i$, strongly simplifying the three-body problem  (more details can be found  in Ref.~\cite{BC12}). The eikonal incoming  distorted wave is given by 
\begin{align}
\label{eq:39}
\mathcal{F}\approx S_{NT}(\ve R_{NT})\,S_{cT}(\ve R_{cT})\,\chi_i(\ve R_{NT},\ve R_{cT}).
\end{align}
where $S_{NT}$ and $S_{cT}$ are respectively the $N$-$T$ and $c$-$T$ eikonal $S$-matrices. The eikonal model exhibit cylindrical symmetry, it is therefore often expressed in terms of the transverse $(b_{(N,c)T})$ and longitudinal $(Z_{(N,c)T})$ coordinates  
\begin{align}
\label{eq:40}
R_{(N,c)T}^2=b_{(N,c)T}^2+Z_{(N,c)T}^2.
\end{align}
In these coordinates, the eikonal $S$-matrices can then be written as
\begin{align}
\label{eq:41}
 &S_{NT}(Z_{NT},b_{NT})=\exp\Big[i\delta^{\rm eik}_{NT}(Z_{NT},b_{NT})\Big],\\
&S_{cT}(Z_{cT},b_{cT})=\exp\Big[i\delta^{\rm eik}_{cT}(Z_{cT},b_{cT})\Big],
\end{align}
in terms of the eikonal phases $\delta^{\rm eik}_{NT}$ and $\delta^{\rm eik}_{cT}$
\begin{align}
\label{eq:43}
 &\delta^{\rm eik}_{NT}(Z_{NT},b_{NT})=-\frac{\mu_{PT}}{\hbar^2 k_{PT}}\int_{-\infty}^{Z_{NT}}U_{NT}(Z'_{NT},b_{NT})\,dZ'_{NT},\\
&\delta^{\rm eik}_{cT}(Z_{cT},b_{cT})=-\frac{\mu_{PT}}{\hbar^2 k_{PT}}\int_{-\infty}^{Z_{cT}}U_{cT}(Z'_{cT},b_{cT})\,dZ'_{cT},\label{eq:44}
\end{align}
with $\hat U_{NT}$ and $\hat U_{cT}$ the $N$-$T$ and $c$-$T$ optical potentials.

By assuming that the nucleon and the core have the same initial velocity as the projectile, i.e., 
\begin{align}
\label{eq:44b}
k_{NT}=\frac{\mu_{NT}}{\mu_{PT}}k_{PT};\quad k_{cT}=\frac{\mu_{cT}}{\mu_{PT}}k_{PT},
\end{align}
the $\chi_i$ eikonal  scattering wave is  simply given by incoming plane waves
\begin{align}
\label{eq:45}
\chi_i(\ve R_{NT},\ve R_{cT})=\exp\Big[i\left(k_{NT}Z_{NT}+k_{cT}Z_{cT}\right)\Big].
\end{align}
In the eikonal model, the prior potential $V_{prior}$ is chosen as the sum of the $c$-$T$ and $n$-$T$ interactions~\cite{HM85,GRetal21}
\begin{equation}
\hat V_{prior}=\hat V_{cT}+\hat V_{NT}.
\end{equation}
As discussed in Ref.~\cite{GRetal21}, this allows to take into account breakup effects in the entrance channel.

In order to further simplify the calculation, one can also approximate the $N$-$T$ and $c$-$T$ final wave functions within the eikonal model
\begin{align}
\chi^{(f_{NT})*}_{NT}(\ve R_{NT})&=S_{NT}^{f*}(\ve R_{NT})e^{-i\ve k^f_{{NT}}\ve R_{NT}},\\
\chi^{(f_{cT})*}_{cT}(\ve R_{cT})&=S_{cT}^{f*}(\ve R_{cT})e^{-i\ve k^f_{{cT}}\ve R_{cT}}.
\end{align}
where $S_{{(N,c)}T}^{f*}$ is the eikonal $S$-matrix and $\ve k^f_{(N,c)T}$ are the $(N,c)$-$T$ final wave vectors.

Then, by approaching the integral over the target degrees of freedom $\ve \xi_T$ in Eq. (\ref{eq:20})  in terms of an eikonal $N$-$T$ final wave function,
 $U_{NT}$ and  $U_{cT}$, the   transition amplitude reads
\begin{align}
  \label{eq:204}
  \nonumber T^{(f_{NT})}(\ve R_{NT},\ve R_{cT}) &\approx S_{NT}(\ve R_{NT})S_{{N}T}^{f*}(\ve R_{NT})\,S_{cT}(\ve R_{cT})\,\\
  &\hspace{-3cm} \times \, [U_{NT}(\ve R_{NT})+U_{cT}(\ve R_{cT})] e^{-i\ve q_{NT}(\ve R_{NT}-Z_{NT}\hat Z)}e^{ik_{cT}Z_{cT}}
  ,
\end{align}
and the source term to be used  within this eikonal framework is
\begin{align}
\label{eq:205}
\nonumber \rho_h^{(f_{NT})}(\ve r)&\approx\\
\nonumber&\hspace{-1.4cm} \int S_{NT}(\ve R_{NT})S_{{N}T}^{f*}(\ve R_{NT})\,S_{cT}(\ve R_{cT})S_{cT}^{f*}(\ve R_{cT})\,  \\
&\hspace{-1.4cm}\times [U_{NT}(\ve R_{NT})+U_{cT}(\ve R_{cT})] \nonumber\\
&\hspace{-1.4cm}\times e^{-i\ve q_{NT}(\ve R_{NT}-Z_{NT}\hat Z)}e^{-i\ve q_{cT}(\ve R_{cT}-Z_{cT}\hat Z)}d\ve R_{cT}
\end{align}
where we define  the transferred momenta $\ve q_{cT}=\ve k^f_{{cT}}-k_{cT}\hat Z$ and $\ve q_{NT}=\ve k^f_{{NT}}-k_{NT}\hat Z$. Note that this expression can be further simplifying assuming $k^f_{(c,N)T}\approx k_{(c,N)T}$, neglecting the dynamics of the reaction, as done in the usual  eikonal model.

Since the wave function (\ref{eq:39}) takes into account  breakup effects \cite{GRetal21},  the eikonal approximation is able to describe physical process in which the  hole or  nucleon absorption  takes place before as well as after (or simultaneously to) the breakup process. However, it should only be applied when the kinematical conditions are suitable for an eikonal approximation, i.e., when the bombarding energy is large enough. It is important to note that, contrary to the usual eikonal description of knockout reactions \cite{HT03,HM85}, the cross section~\eqref{eq:33} accounts for non-sudden effects in the breakup of the projectile, by treating explicitly the dynamics of the core-hole system in terms of the hole optical potential $\hat U_h$.

Finally, let us stress that the extension of the eikonal approximation to treat explictly non-local $N$-$T$ and $c$-$T$ optical potentials is not straighforward \cite{Hebborn:21}. The issue  lies in the fact that non-local interactions depend on a  integral over the whole space of the  wave function and the non-local potential, while the eikonal wave function is not accurate at short distances.   One way to avoid this issue is to derive the  local-equivalent potentials, (i.e, local potentials producing the same elastic phase shifts as the original non-local ones), and to use them to compute the eikonal phase shifts~\eqref{eq:43}--\eqref{eq:44}.

\subsection{Distorted Wave Born Approximation}\label{DWBA}

In what we will call the DWBA approximation to the GFK formalism, the distorted wave $\mathcal{F}$ is approximated by 
\begin{align}
  \label{eq:206}
\mathcal{F}  \approx \chi_{PT}(\ve R),
\end{align}
where $\chi_{PT}$ is the solution of the \Sch equation with the  projectile-target optical potential $\hat U_{PT}$
\begin{align}
\label{eq:206b}
  \left(\hat T_{PT}+\hat U_{PT}-E\right)\chi_{PT}(\ve R)=0.
\end{align}
The potential $\hat V_{prior}$ will now include the remnant term associated with the standard DWBA,
\begin{align}
\hat V_{prior}=\hat V_{cT}+\hat V_{NT} -\hat U_{PT}.
\end{align}

In a similar spirit as the one we adopted in the eikonal approach, one can also approximate the $N$-$T$ and  $c$-$T$ final wave functions $\chi^{(f_{NT})}_{NT}$ and $\chi^{(f_{cT})}_{cT}$ by the  solutions of the \Sch equations with the optical potentials $U_{NT}$ and $U_{cT}$, respectively.   
The integral over the target degrees of freedom $\ve \xi_T$ in Eq. (\ref{eq:20})  can then be evaluated in terms of a $N$-$T$ final wave function,
$U_{NT}$ and  $U_{cT}$
\begin{align}
\label{eq:207}
\nonumber T^{(f_{NT})}&(\ve R_{NT},\ve R_{cT})\approx \chi_{PT}(\ve R)\chi^{(f_{NT})*}_{NT}(\ve R_{NT})\\
&\times\left[U_{NT}(\ve R_{NT})+U_{cT}(\ve R_{cT})-U_{PT}(\ve R)\right],
\end{align}
and the source term becomes
\begin{align}
\label{eq:208}
\nonumber \rho_h^{(f_{NT})}&(\ve r)\approx\int \chi^{(f_{cT})*}_{cT}(\ve R_{cT})\chi_{PT}(\ve R)\chi^{(f_{NT})*}_{NT}(\ve R_{NT})\\
&\hspace{-0.5cm}\times\left[U_{NT}(\ve R_{NT})+U_{cT}(\ve R_{cT})-U_{PT}(\ve R)\right]d\ve R_{cT}.
\end{align}

Let us  emphasize that  $\chi^{(f_{NT})}_{NT}$ can be calculated with the help of the optical reduction of the Green's function $\hat G_{NT}$~\eqref{eq:8},
 \begin{align}
\label{eq:36}
&\nonumber  \hat  G^{opt}_{NT}(E_{f_{NT}})\nonumber \\
&\nonumber= \lim_{\eta\to0}\left\langle \phi_T^{(0)}\right|\left(E_{f_{NT}}-\hat T_{NT}-\hat V_{NT}-\hat h_T+i\eta\right)^{-1}\left|\phi_T^{(0)}\right\rangle\\
&=\lim_{\eta\to0}\left(E_{f_{NT}}-\hat T_{NT}-\hat U_{NT}-\epsilon_T^{(0)}+i\eta\right)^{-1},
\end{align}
where  $\epsilon_T^{(0)}$ is the  target ground-state energy.
This single-particle Green's function can be calculated numerically with Lagrange mesh techniques~\cite{DB10}. Although straighforward for local potentials, computing  Green's functions for non-local potentials  is  not trivial and will be reported in another  contribution~\cite{HPprep}. The overlap can then be obtained making use of the relation 
\begin{eqnarray}
\label{eq:37}
&&\chi_{NT}^{(f_{NT})*}(\ve r;E_{f_{NT}})\chi_{NT}^{(f_{NT})}(\ve r;E_{f_{NT}})=\nonumber\\&&\hspace{3cm}-\frac{1}{\pi}\,\text{Im }  G^{opt}_{NT}(\ve r,\ve r;E_{f_{NT}}).
\end{eqnarray}
As mentioned in Sec.~\ref{SecGFK}, this procedure enforces the proper normalization of the overlap as  the spectroscopic factor is directly encoded in the energy dependence of the Green's function [see Eq.~\eqref{eq:19}]. Since the cross section   (\ref{eq:33}) is a functional of $\chi_{NT}^{(f_{NT})*}\chi_{NT}^{(f_{NT})}$, it is unchanged under an arbitrary phase change $\chi_{NT}^{(f_{NT})}\to e^{i\varphi}\chi_{NT}^{(f_{NT})}$, and the expression~\eqref{eq:37} is enough to provide   $\chi_{NT}^{(f_{NT})}$. 
Once the overlap has been determined, the source term (\ref{eq:208}) and the hole wave function (\ref{eq:30})
can be computed by numerical integration.

As for the eikonal model, the DWBA formulation of the GFK  accounts for  processes in which breakup has been induced. The main difference here is that the approximation is valid for low bombarding energies, for which the eikonal approximation may not be accurate.

\section{Conclusions}\label{Conclusions}

 One-nucleon knockout and  transfer reactions are key  probes of the single-particle structure of nuclei away from stability. The standard theoretical approach  associated with these observables  rely on  spectroscopic factors derived within some nuclear   structure formalism, and reaction cross sections, and  the discrepancy between theory and experiment is often associated  with  missing correlations in the  structure description \cite{Aetal21}. A striking feature of the comparison between the theoretical  and  experimental knockout observables  is  a  marked neutron-to-proton asymmetry dependence which is not observed in the analysis of transfer and quasifree reactions  \cite{Aetal21,Gade:08,TG14,TG21,GM18,Letal10,Tetal09}. In order   to understand what causes this discrepancy, it is pressing to describe both of these reaction processes within the same  framework, providing a unified description of  structure and reaction.

 In this work, we introduce a new formalism, the GFK, which describes one-neutron knockout and transfer reactions making use of dispersive optical potentials, hence treating on the same footing bound and scattering states.   
 For one-nucleon addition transfer reactions, which are typically measured at low to medium energies (from few MeVs to 50~MeV/nucleon), the use of a DWBA incoming scattering function leads to the GFT formalism~ \cite{PotelNunesThompson}. Moreover, the GFK can also predict one-nucleon removal reactions, such as $(p,d)$ and knockout reactions, thus allowing the description of transfer and knockout reactions within the same framework.
 
 Because the GFK relies on Green's functions, the link between the few-body problem and the underlying nuclear structure of the ground states of the target and the projectile is made explicit.  Moreover, no core spectator or sudden approximation is made, which allows to include dynamical effects associated with the nucleon  extraction  from the projectile,  such as  excitation of the core above the particle emission threshold during the collision.  Our analysis suggests that the  discrepancy observed in the analysis of knockout and  transfer data might arise from these dynamical effects, that are  neglected in the usual eikonal model. 
 
The main approximations made in the GFK {are the assumptions that the prior potential does not depend on the intrinsic coordinates of the core and} the factorization of the many-body wave function \eqref{eq:200}, in which the effect of the projectile-target interaction is described by a incoming scattering function $\mathcal{F}$. The choice of this  function reflects the approximation of the few-body problem we are willing to make, and can be adapted to a specific energy regime. In particular, {we discuss an  eikonal  and a DWBA approximation}. We plan to test the validity of these approximations and verify their applicability for different systems, i.e., with various beam energies and for nuclei ranging from  the valley of stability to the proton and neutron driplines.

In a future publication we plan to compare knockout and transfer observables obtained within a standard reaction model, with the  GFK calculation along an isotopic chain. For this, we plan to use the  dispersive optical model developed in Refs. \cite{Dickhoff:19,Petal20PRC,Petal20PRL,AD19}, which  provides a description of structure and reaction properties for  nuclei exhibiting different neutron-to-proton asymmetry. This study will provide quantitative estimates of the dynamical effects associated with the extraction of the nucleon, and might help to explain  the systematic discrepancy observed by Gade \etal~ \cite{Gade:08,TG14,TG21}. 

\begin{acknowledgements}
	The authors thank P. Capel, W. H. Dickhoff, J. E. Escher, T. Frederico and C. D. Pruitt for insightful discussions and J. E. Escher and C. D. Pruitt for their careful rereading of this manuscript. 
C.~H. acknowledges the support of  the U.S. Department of Energy, Office of Science, Office of Nuclear Physics, under the FRIB Theory Alliance award no. DE-SC0013617 and  under Work Proposal no. SCW0498.  This work was performed under the auspices of the U.S. Department of Energy by Lawrence Livermore National Laboratory under Contract No. DE-AC52-07NA27344.
\end{acknowledgements}

\bibliographystyle{apsrev}

\end{document}